\documentclass[11pt,paper,letterpaper]{JHEP3}
\usepackage{epsfig}
\usepackage{psfrag}
\usepackage{amsmath}
\newcommand{\bq}{\begin{eqnarray}}
\newcommand{\eq}{\end{eqnarray}}
\newcommand{\bqs}{\begin{eqnarray*}}
\newcommand{\eqs}{\end{eqnarray*}}
\newcommand{\p}{\partial}

\def\L{\Lambda}

\def\d{\delta}
\def\e{\varepsilon}
\def\vp{\varphi}
\def\g{\gamma}
\def\k{\kappa}   
\def\l{\lambda}
\def\m{\mu}
\def\n{\nu}
\def\o{\omega}                    
\def\th{\theta}                                   
\def\s{\sigma}                
\def\O{\Omega}
\title{Five-dimensional metrics of Petrov type 22}
\author{Pieter-Jan De Smet\\
C.N. Yang Institute of Theoretical Physics\\
State University of New York\\
Stony Brook, NY 11794-3840, USA\\
E-mail: {\tt Pieterj@insti.physics.sunysb.edu} }
\preprint{YITP-SB-03/7 \\{\tt gr-qc/0302081}}
\abstract{We classify all five-dimensional Einstein manifolds that are
static, have an $SO(3)$ isometry group and have Petrov type~22. We use this
classification to show that the localized black hole in the Randall--Sundrum
scenario necessarily has Petrov type~4.}
\begin{document}
\section{Introduction}
In this article, we study the metric of a black hole that is localized on a 
brane in the five-dimensional AdS-space. This metric is of interest because it 
presumably describes the gravitational collapse 
of matter trapped on a brane in the Randall--Sundrum 
scenario~\cite{RS}. It was argued in ref.~\cite{9909205} that this black hole is 
described by a ``black cigar'' in five dimensions. 
However the analytical form of the black cigar 
metric is not known. For an attempt to find this metric using a restricted
ansatz, see ref.~\cite{0110298}. Some numerical approximations can be found in 
refs.~\cite{Kudoh,Wiseman}. The analytical form of the metric of 
a black hole living on a membrane in four-dimensional 
AdS-space was constructed in ref.~\cite{9911043}.  

We use the five-dimensional Petrov 
classification~\cite{0206106} to look for this metric. We show that this 
unknown metric should have Petrov type~22 or type~4. In the latter case, we do
not have additional constraints on the Weyl tensor. Imposing Petrov type~22, 
however, leads to an additional constraint on the Weyl tensor. This makes it
easier to solve Einstein's equations. We therefore look for the metric of the 
localized black hole within the class of metrics having Petrov type~22. This 
method is similar to the method used by Kerr for constructing the rotating black
hole in four dimensions. This metric was found while sieving through the metrics
of Petrov type~$D$.

The outline of this article is as follows. In Section~\ref{s:Petrov}, we give 
a short review of the five-dimensional Petrov classification. There, we also 
show that the Petrov type of a black hole that is localized on a brane should have
Petrov type~4 or~22. In Section~\ref{s:22}, we 
classify all metrics that have Petrov type~22, are 
static\footnote{although it was argued in ref.~\cite{0108013} 
that the localized black hole should be time-dependent. See 
ref.~\cite{0206155} for an AdS/CFT argument.} and have a space-like 
$SO(3)$ isometry group. In Section~\ref{s:22s}, we classify all metrics that have
Petrov type~\underline{22}, are static and have a space-like 
$SO(3)$ isometry group. From these classifications, it follows that the black hole
we were looking for, does not have type~22. Hence, unfortunately, we did
not find the analytical form of its metric.
\section{Review of the five-dimensional
Petrov classification}\label{s:Petrov}
We only give a brief review of this classification, a longer discussion can 
be found in~\cite{0206106}. We need to introduce two objects, the 
\textit{Weyl spinor} and the \textit{Weyl polynomial}. 

The Weyl spinor $\Psi_{abcd}$
is the spinorial translation of the Weyl tensor $C_{ijkl}$
$$\Psi_{abcd} = (\g_{ij})_{ab} (\g_{kl})_{cd}C^{ijkl}.$$
Here, $\g_{ij} = \frac{1}{2} [\g_i,\g_j ]$, where $\g_i$ are
the $\g$-matrices in 5 dimensions. These are $4\times 4$ matrices. It can be 
verified that the Weyl spinor is symmetric in all its indices.

The Weyl polynomial $\Psi$ is a homogeneous polynomial of degree four in four variables:
\begin{equation}
\Psi = \Psi_{abcd} x^a x^b x^c x^d\ .
\end{equation}
The Petrov type of a given Weyl tensor is nothing else than the number and
multiplicity of the irreducible
factors of $\Psi$. 
In this way, we get 12 different
Petrov types, which are depicted in figure~\ref{fig:Ptypes}.
\FIGURE[ht]{
\begin{psfrags}
\psfrag{1}[][]{4}
\psfrag{2}[][]{31}
\psfrag{3}[][]{22}
\psfrag{4}[][]{211}
\psfrag{5}[][]{\underline{22}}
\psfrag{6}[][]{2\underline{11}}
\psfrag{7}[][]{1111}
\psfrag{8}[][]{\underline{11} \underline{11}}
\psfrag{9}[][]{11\underline{11}}
\psfrag{10}[][]{$\Psi = 0$}
\psfrag{11}[][]{\underline{1111}}
\psfrag{12}[][]{1\underline{111}}
\epsfig{file=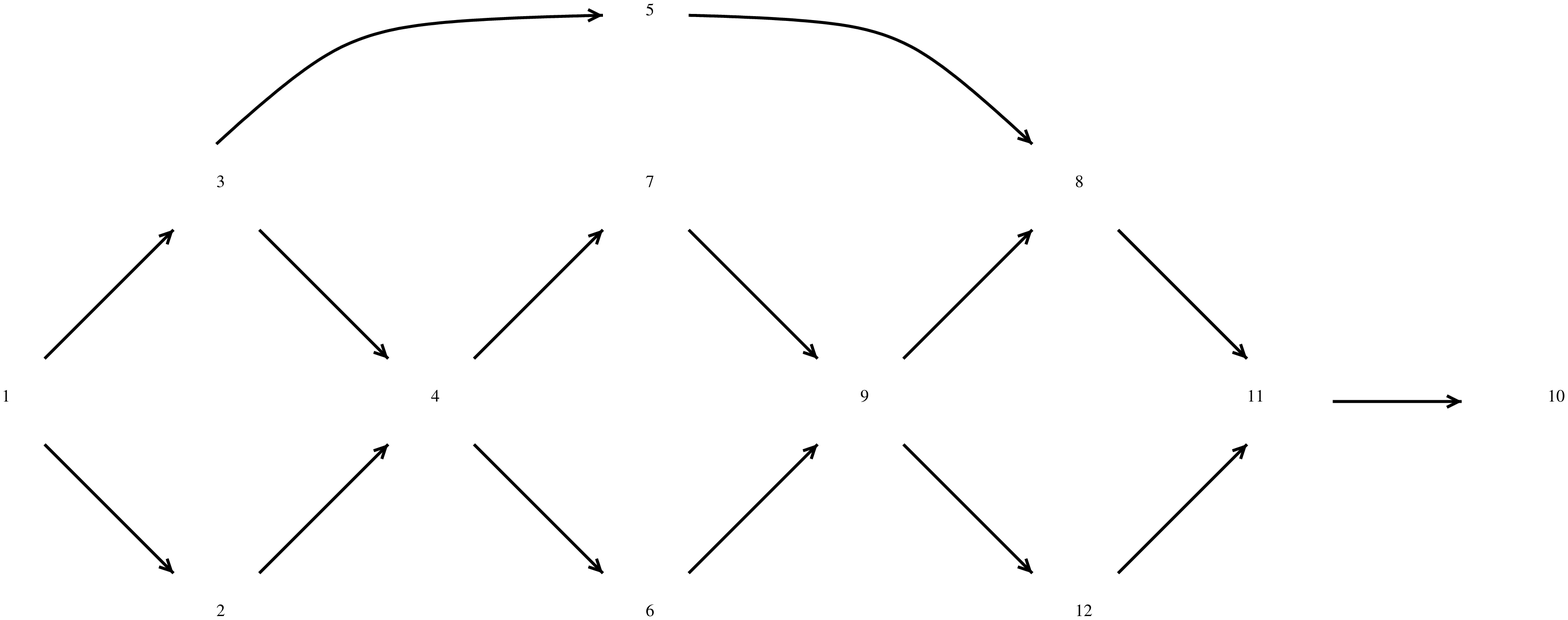,width=10cm,height=5cm}
\end{psfrags} 
\caption{The 12 different Petrov types in 5 dimensions.}
\label{fig:Ptypes}
}
We use the following notation. The number denotes the degree of the
irreducible factors and underbars denote the multiplicities. For example, a Weyl
polynomial which can be factorized into two different factors, each having
degree~2, has Petrov type~22. As a second example, 
a Weyl polynomial which can be
factorized in a polynomial of degree~2 times the square of a polynomial of
degree~1 has Petrov type~2\underline{11}. The arrows between the different
Petrov types denote increasing specialization of the Weyl tensor. 
In accordance to the literature on the four-dimensional Petrov classification, 
all Weyl tensors different from type~4 are called algebraically special. 
\subsubsection*{Example: the black string in AdS}
The metric of a black string in AdS reads~\cite{9909205}
\begin{equation}\label{BS}
ds^2 = \frac{l^2}{z^2}\left(U dt^2+U^{-1} dr^2 + dz^2 + r^2 d\O_2^2 \right),
\end{equation}
where $U(r) = 1 - \frac{2 m}{r}$ and $d\O_2^2$ is
the metric on the unit 2-sphere.
If we choose the tetrad as 
\begin{alignat*}{3}
e_t &= \frac{z}{l} U^{-1/2} \p_t\qquad &e_r &= \frac{z}{l} U^{1/2} \p_r\qquad
&e_{z} &= \frac{z}{l}\p_z\\
e_{\theta} &= \frac{z}{l r}\p_{\theta}\qquad& 
e_{\phi} &= \frac{z}{l r \sin\th}\p_{\phi},&&
\end{alignat*}
we find after some algebra the Weyl polynomial
$$\Psi = -\frac{48 m z^2}{l^2 r^3}
\left(T^2 X^2 + X^2 Y^2 + T^2 Z^2 +Y^2 Z^2 - 4 T X Y Z\right).$$
The metric~\eqref{BS} has Petrov type $22$ because
$\Psi$ can be factorized into two polynomials of degree~2
$$\Psi = -\frac{48 m z^2}{l^2 r^3} 
\left( T X + i XY - i T Z - YZ\right)\left( T X - i XY +i T Z - YZ\right).$$
Far from the AdS-horizon, the localized black hole will look like this black
string~\cite{9909205}. Because the black string has type~22, it follows from 
Figure~\ref{fig:Ptypes} that the localized black hole should have Petrov 
type~22 or~4. In the next Section, we will look for the metric of the localized
black hole by classifying the metrics of type~22.
\section{Classification of metrics of type~22}\label{s:22}
In this section, we classify all metrics that are static, have an $SO(3)$-isometry 
group and have Petrov type~22. These metrics satisfy Einstein's equations. Hence,
the Ricci-tensor satisfies $R_{\mu\nu} = 4 \L g_{\mu\nu}$. We will assume in the
rest of the article that $\L\neq0$. The classification in the case of $\L=0$ can 
be found in ref.~\cite{0206106}.

A calculation shows that the most general ansatz for the Weyl spinor
$\Psi_{abcd}$ has 4 independent components.
\begin{equation}\label{Wans}
\begin{split}
&\Psi_{1114} = \Psi_{1444}=\Psi_{2223} = \Psi_{2333}=24 i  \vp_4\\
&\Psi_{1123} = \Psi_{1224}=\Psi_{1334} = \Psi_{2344}=-8 i \vp_4\\
&\Psi_{1111} = \Psi_{2222}=\Psi_{3333} = \Psi_{4444}=-24 ( \vp_2 + \vp_3)\\
&\Psi_{1133}= \Psi_{2244}=8 ( \vp_2 + \vp_3)\\
&\Psi_{2233} = \Psi_{1144}=8 ( \vp_2 -3  \vp_3)\\
&\Psi_{1122} =\Psi_{3344} =8(-2\vp_1-\vp_2+\vp_3)\text{ and }
\Psi_{1234} =8 (\vp_1 + \vp_3)
\end{split}
\end{equation}
Here, the four functions $\vp_i$ depend on the coordinates $r$ 
and $z$. The above linear combinations are chosen in such a way 
that the functions $\vp_i$ transform nicely under tetrad transformations 
(see appendix~\ref{A:transf}).

A metric has Petrov type~22 if by definition the 
Weyl 
polynomial obtained from the Weyl spinor~\eqref{Wans} can be factorized into two
factors of degree~2. In~\cite{0206106}, we have shown that this is the case if and 
only if one of the 
following four conditions is satisfied
\begin{equation}\label{Pcond2}
\begin{aligned}
\text{Case A: }\ &\vp_3=\vp_4=0\\
\text{Case B: }\ &\vp_1+\vp_2=0
\end{aligned}\qquad
\begin{aligned}
\mbox{Case C: }\ &\vp_1^2=\vp_3^2+\vp_4^2\\
\mbox{Case D: }\ &\vp_2^2=\vp_3^2+\vp_4^2
\end{aligned}
\end{equation}
The classification is then as follows. 

\textit{The metric has type~22 if and only if it is one of the following metrics.}
\begin{enumerate}
\item
The black string in AdS~\cite{9909205}
\begin{equation}
ds^2 = \frac{l^2}{z^2}\left(U dt^2+U^{-1} dr^2 + dz^2 + r^2 d\O^2 \right),
\end{equation}
where $U(r) = 1 - \frac{2 m}{r}$ and $d\O^2$ is
the metric on the unit 2-sphere.
\item
A wrapped product built on a dS-Schwarzschild space
\begin{equation}\label{LamkegeldSS}
ds^2 = dz^2 + 
\sin^2(\sqrt{\L}\ z)\left[\left(1-\frac{2 m}{r}-\L r^2 \right) dt^2 +
\frac{dr^2}{1-\frac{2 m}{r}-\L r^2 } + r^2 d\O^2 \right]
\end{equation}
This metric is part of the so called RS-AdS-Schwarzschild black 
hole~\cite{Kaloper,Nihei}
\item
A cosmological model in which space is the product of a two-sphere 
and a two-dimensional manifold of constant curvature. Hence, the metric
reads
\begin{equation}\label{sphere-constant}
ds^2 = dz^2 + a(z)^2 d\O^2+ b(z)^2 dM^2,
\end{equation}
where $M$ can be $\mathbb{E}^2$, $S^2$ or $\mathbb{H}^2$.
The functions $a(z)$ and $b(z)$ are determined by Einstein's equations.
We do not know the analytical expression of the general solution. 
A particular solution is $a(z) =b(z) =\sin(\sqrt{\L}\ z)/\sqrt{3 \L}$. 
See ref.~\cite{Hervik} for a classification of all $4+1$-dimensional
cosmological models with spatial hypersurfaces
that are homogeneous, connected and simply connected.

\end{enumerate}
The derivation of this classification is tedious, it 
can be found in appendices~\ref{ap:ABandC} and~\ref{ap:D}. 
We now prove\footnote{We remind the reader that we have assumed that the 
metric of the localized black hole is static. If the results in ref.~\cite{0108013} and 
ref.~\cite{0206155} are valid, this restriction is too strong and our proof breaks down.}  
that the localized black hole on a brane has Petrov type~4. In 
Section~\ref{s:Petrov}, we argued that the metric of the localized black hole 
has type~22 or type~4. This metric does not appear in our classification of 
metrics of type~22. Hence, it has Petrov type~4. 
\section{Classification of metrics of type~\underline{22}}\label{s:22s}
A metric has Petrov type~\underline{22} if by definition the 
Weyl polynomial is the square of a polynomial of degree~2. 
This is the case if and only if one of the 
following three conditions is satisfied
\begin{equation}
\begin{aligned}
\text{Case AC: }\ &\vp_1=\vp_3=\vp_4=0\\
\text{Case AD: }\ &\vp_2=\vp_3=\vp_4=0\\
\mbox{Case BC: }\ &\vp_1+\vp_2=0 \quad\text{and}\quad \vp_1^2=\vp_3^2+\vp_4^2\\
\end{aligned}
\end{equation}
These conditions are combinations of the
conditions~\eqref{Pcond2}. This is reflected in our notation.
The classification is then as follows. 

\textit{The metric has type~\underline{22} if and only if it is one of 
the following metrics.}
\begin{enumerate}
\item
The product of the spheres $S^3$ and $S^2$
\begin{equation}\label{S2S3}
ds^2 = \frac{1} {2 \L} d\O_3^2 + \frac{1}{4 \L} d\O_2^2
\end{equation}
\item
The generalized AdS-Schwarzschild metric~\cite{Birmingham}
\begin{equation}\label{kLambda}
ds^2 = U(r) dt^2+U^{-1}(r) dr^2 + r^2 dM^2,
\end{equation}
where $U(r) = k - \frac{m}{r^2} + \L r^2$ and $M$ is a space of constant curvature. 
If $M = S^3$, $\mathbb{E}^3$ or $\mathbb{H}^3$, then $k=1$, 0, $-1$ respectively.
If $M$ is the unit three-sphere, then the metric 
reduces to AdS-Schwarzschild. 
\end{enumerate}
Details of the classification can be found in appendix~\ref{ap:22s}.
\section{Conclusions}\label{s:conc}
In this article, we have given a classification of five-dimensional Einstein
manifolds that are static, have an $SO(3)$ isometry group and have Petrov
type~22. We have used this classification to prove that the localized black hole 
on a brane is not algebraically special. Some topics of further research are 
the following.

\begin{enumerate}
\item
One can further refine the five-dimensional Petrov classification 
as given in~\cite{0206106}. Indeed, it should 
be possible to subdivide the Petrov types for which the Weyl polynomial can not be 
completely factorized into smaller subtypes
(i.e.~the Petrov types 4, 31, 22, \underline{22} and 2\underline{11}). 
This project is essentially the
determination of an invariant classification of the polynomials of degree four 
in four variables, hence belongs to 19th century's ``invariant theory''. See 
ref.~\cite{Olver} for an introduction to classical invariant theory.
\item 
Classify all Einstein metrics of type~$22$ without assuming additional isometries. 
This classification is a higher-dimensional
generalization of Kinnersley's classification of four-dimensional
metrics of Petrov type~$D$.
This could lead to the discovery of new five-dimensional metrics. It would shed
some light on the possible solutions of the five-dimensional Einstein's equations.
One knows that
the set of black holes is much richer in five dimensions than in four
dimensions. See e.g.~ref.~\cite{ring} for a five-dimensional rotating black string.

Before attacking the general classification, it is perhaps better
in view of refs.~\cite{0108013} and~\cite{0206155}
to start by classifying 
time-dependent metrics with an $SO(3)$ isometry group.
\item
As far as I know, almost nothing is known about the five-dimensional Petrov 
classification. It would be good to make a thorough study of it.
\end{enumerate}
\section*{Acknowledgments}
This work has been supported in part by the NSF grant PHY-0098527.
\appendix
\section{Tetrad transformation on the Weyl spinor}\label{A:transf}
Under tetrad rotations ($s = +1$) and reflections ($s = -1$) 
$$
\begin{pmatrix}
e_r\\e_z
\end{pmatrix} \to 
\begin{pmatrix}
\cos\chi & \sin\chi\\
-s\sin\chi & s\cos\chi\\
\end{pmatrix}
\begin{pmatrix}
e_r\\e_z
\end{pmatrix},
$$
the Weyl spinor transforms as
$$
\vp_1\to\vp_1,\quad\vp_2\to\vp_2,\quad
\begin{pmatrix}
\vp_3\\ \vp_4
\end{pmatrix} \to 
\begin{pmatrix}
\cos 2\chi & -\sin 2\chi\\
s\sin 2\chi &s\cos 2\chi\\
\end{pmatrix}
\begin{pmatrix}
\vp_3\\ \vp_4
\end{pmatrix}.
$$
From this, we have the important result that we can always do a tetrad 
rotation to put $\vp_4=0$. On top of this, we can specify 
the sign of $\vp_3$ by an additional reflection. 
\section{The field equations}\label{veldvgl}
The vacuum Einstein equations (notice our non-standard normalization of the 
cosmological constant)
\begin{equation}\label{Einstein}
R_{\mu\nu} - \frac{1}{2} g_{\mu\nu} R -14 \L g_{\mu\nu}=0
\end{equation}
are equivalent to the system
\bq
&&d\o +\o \wedge \o=\O\label{NPb}\\
&&d\O + \o \wedge\O -\O\wedge\o =0\label{NPc},
\eq
where the Riemann 
tensor $\O$ is expressed in terms of the Weyl spinor $\Psi_{abcd}$~\eqref{Wans}
and the Ricci tensor, which
is proportional to the metric by~\eqref{Einstein}. The connection~$\o$ is the Levi-Civita
connection associated with the tetrad. 
The most general ansatz for the tetrad of a static space-time with $SO(3)$
symmetry reads
\begin{equation}\label{eans}
\begin{split}
&e_t = A\ \p_t\\ 
&e_r\mbox{ and }e_z:\mbox{ linear combinations of }\p_r 
\mbox{ and }\p_z\\ 
&e_{\th} = \s\ \p_{\theta}\\
&e_{\phi} = \frac{\s}{\sin\th}\ \p_{\phi}
\end{split}
\end{equation}
In the above, $A$ and $\s$ are functions of the coordinates $r$ and $z$. 
The non-zero commutators are
\begin{subequations}
\label{eq:comms}
\bq
\left[e_t,e_r\right]&=& \m e_t,
\qquad\left[e_t,e_z\right]= \n e_t,\label{comm:et}\\
\left[e_r,e_{\th}\right]&=& \k e_{\th},\qquad [e_z,e_{\th}]= \l e_{\th},\\
\left[e_r,e_{\phi}\right]&=& \k e_{\phi},\qquad [e_z,e_{\phi}]= \l e_{\phi},\\
\left[e_r,e_z\right]&=& \d e_r+\e e_z\label{comm:erea}\\
\left[e_{\th},e_{\phi}\right]&=&- \cot\th\ \s\ e_{\phi},
\eq
\end{subequations}
where $\m,\n,\d,\e,\k,\l$ and $\s$ are functions of the
coordinates $r$ and $z$.
We will split the resulting field equations into two blocks. 
\subsection*{The equations from~\eqref{NPb}}
The definition of the curvature leads to two algebraic equations 
\bq
&&\s^2-\k^2-\l^2-\L -3\vp_1-\vp_2=0\label{eq:algs}\\
&&\k\m+\l\n -\L + \vp_1-\vp_2=0\label{eq:alg}
\eq
\begin{subequations}
\label{eq:curvr}
and a set of first order differential equations. The ones 
involving the vector $e_r$ are
\bq
e_r(\s) &=& \s\k\\
e_r(\e) -e_z(\d)&=& \d^2 +\e^2  +  \vp_1 + 3\vp_2+\L\\
e_r(\k) &=& \k^2 - \d\l - \vp_1 - \vp_2 -2\vp_3+\L\\
e_r(\l) &=& \k\d + \k\l + 2\vp_4\label{eq:erl}\\
e_r(\m) &=& -\n\d -\m^2-\vp_1+\vp_2-4\vp_3-\L\label{eq:erm}\\
e_r(\n) &=& -\n\m + \d\m + 4\vp_4\label{eq:ern}
\eq
\end{subequations}
and the ones involving the vector $e_z$ are
\begin{subequations}
\label{eq:curva}
\bq
e_z(\s) &=& \s\l\\
e_z(\k) &=& -\e\l +\k\l +2\vp_4\label{eq:eak}\\
e_z(\l) &=&\e\k +\l^2 -\vp_1 -\vp_2 +2\vp_3+\L\label{eq:eal}\\
e_z(\m) &=& -\n\e -\n\m +4\vp_4\label{eq:eam}\\
e_z(\n) &=&-\n^2 +\e\m -\vp_1 +\vp_2 +4\vp_3-\L\label{eq:ean}
\eq
\end{subequations}
\subsection*{The equations from~\eqref{NPc}}
The Bianchi identity leads the following set of equations
\begin{equation}
\label{eq:Bianchi}
\begin{split}
e_r(\vp_1) &=
     \tfrac{5}{2}\k\ \vp_1 + 
       \tfrac{1}{2}\ (\k + \m)\vp_2 + 
       \tfrac{1}{2}\ (4\k + \m)\vp_3 - 
       \tfrac{1}{2}\ (\n + 4\l)\vp_4\\ 
e_r(\vp_2) &=
     \tfrac{1}{2}\k\ \vp_1 + 
       \tfrac{1}{2}\ (5\k - 3\m)\vp_2 - 
       \tfrac{1}{2}\ (4\k + 3\m)\vp_3 + 
       \tfrac{1}{2}\ (3\n + 4\l)\vp_4\\ 
e_r(\vp_3) &=
      e_z(\vp_4) + 
       \tfrac{1}{2}\k\vp_1-\tfrac{1}{2}\ (\k +\m)
	\vp_2 + (2\ \e +\k - \tfrac{1}{2}\ \m)
	\vp_3\\ 
	& \qquad+ (\tfrac{1}{2}\ \n + 2\d- \l)\vp_4\\ 
e_r(\vp_4) &=-e_z(\vp_3) - 
       \tfrac{1}{2}\ \l\ \vp_1 + \tfrac{1}{2}\ (\n + \l) 
	\vp_2 + (-\tfrac{1}{2}\ \n - 
              2\d+ \l)\vp_3\\
	 &\qquad + (2\ \e +\k - \tfrac{1}{2}\ \m)\vp_4\\ 
e_z(\vp_1) &=
     \tfrac{5}{2}\ \l\ \vp_1 + 
       \tfrac{1}{2}\ (\n + \l)\vp_2 - 
       \tfrac{1}{2}\ (\n + 4\l)\vp_3 - 
       \tfrac{1}{2}\ (4\k +\m)\vp_4\\ 
e_z(\vp_2) &=
     \tfrac{1}{2}\ \l\vp_1 + 
       \tfrac{1}{2}\ (-3\n + 5\l)\vp_2 + 
       \tfrac{1}{2}\ (3\n + 4\l)\vp_3 + 
       \tfrac{1}{2}\ (4\k + 3\m)\vp_4
\end{split}
\end{equation}
\section{Type \underline{22}: solutions in cases AC, AD and BC}
\label{ap:22s}
In this appendix, we classify all solutions of the field equations given in
appendix~\ref{veldvgl} that are of Petrov type~\underline{22}. We point out that
if all components of the Weyl spinor~\eqref{Wans} vanish, the space has constant 
curvature. Indeed, a conformally flat metric which is Einstein, is necessarily of 
constant curvature.
\boldmath
\subsection*{Case AC: $\vp_1=\vp_3=\vp_4=0$}
\unboldmath
From the Bianchi equations~\eqref{eq:Bianchi}, we immediately get
$(\k+\m)\vp_2=0$ and $(\l+\n)\vp_2=0$.
Hence, we have two possibilities. The first one is $\vp_2=0$. Then all 
components of the Weyl spinor are zero and we obtain a space of constant curvature. 
The second possibility is $\k+\m=0$ and $\l+\n=0$. Then we obtain 
from~\eqref{eq:algs} and~\eqref{eq:alg} $\s=0$. This is not a
good solution of the field equations, because in this case the
tetrad~\eqref{eans} is degenerate. Therefore, in case~AC, we obtain only 
the five-sphere.
\boldmath
\subsection*{Case AD: $\vp_2=\vp_3=\vp_4=0$}
\unboldmath
This case is similar to the previous one.
From the Bianchi equations~\eqref{eq:Bianchi}, we get $\k\vp_1 = 0$ and 
$\l\vp_1=0$. If $\vp_1=0$, we have a sphere. 
If $\k=\l=0$, we find from equations~\eqref{eq:alg} and~\eqref{eq:algs} 
$\s^2 = 4 \L$. Hence, the metric reads 
$$ds^2 = dM_3^2 + \frac{1}{4 \L} d\O_2^2,$$
where $\O_2$ is the unit two-sphere. This metric is Einstein if and only if 
$M_3$ is an Einstein space. As is well-known --- see for example
~\cite[Ex. 28.2]{Eisenhart}--- a three-dimensional Einstein space is necessarily
of constant curvature. Therefore $M_3$ is a sphere. All in all, Case AD leads to
the five-sphere or to the product of two spheres~\eqref{S2S3}.
\boldmath
\subsection*{Case BC: $\vp_1+\vp_2=0$ and $\vp_1^2=\vp_3^2+\vp_4^2$}
\unboldmath
As shown in appendix~\ref{A:transf}, we can choose our tetrad in such a 
way that $\vp_4=0$ and
$\vp_3=\vp_1$. We will assume that $\vp_1\neq0$ otherwise, we obtain the 
five-sphere. 
From the Bianchi equations~\eqref{eq:Bianchi}, we obtain 
$\n\vp_1 =\d\vp_1=(\k-\e)\vp_1=0$.
Hence, we have $\n=\d=\k-\e=0$. 
From equation~\eqref{eq:alg}, we get $\vp_1=-1/2\ \k\m + 1/2\ \L$.
Inserting all this into equation~\eqref{eq:curvr} and~\eqref{eq:curva}
leads to the following equations\\
\begin{subequations}
\label{BC:curv}
\begin{minipage}{7.5cm}
\bq
&&e_r(\k) =\k^2+\k\m\\
&&e_r(\l)=\k\l\\
&&e_r(\m)= 3\k\m-\m^2- 4 \L
\eq
\end{minipage}
\begin{minipage}{7.5cm}
\bq
&&e_z(\k) =0\\
&&e_z(\l) =\k^2+\l^2 -\k\m+2 \L\\
&&e_z(\m) = 0
\eq
\end{minipage}
\end{subequations}
\\[2ex]
We will resolve two cases $\k\neq 0$ and $\k = 0$.
\begin{itemize}
\item\underline{Case BC.1: $\k\neq0$}\\
From the commutation relation~\eqref{comm:erea}, we see that we can always find a 
coordinate system in which
$e_r= -r\k \p_r$ and $e_z = 1/r \p_z$. 
The solution of the differential equations~\eqref{BC:curv} then reads
$$
\k=\left(\frac{C_1}{r^4}+\frac{C_2}{r^2} - \L \right)^{1/2},\quad
\l= - \frac{\sqrt{C_2}}{r} \cot\left(\sqrt{C_2} z \right), 
$$ and
$$
\m=\left( \frac{C_1}{r^4} + \L\right) 
\left(\frac{C_1}{r^4}+\frac{C_2}{r^2} - \L \right)^{-1/2}.
$$
Here $C_1$ and $C_2$ are two arbitrary integration constants. We have used a
shift on the coordinate~$z$ to eliminate the third integration constant. 
Finally, the remaining tetrad vectors $e_t$ and $e_\th$ are 
determined from the commutation relation~\eqref{comm:et} and the 
algebraic equation~\eqref{eq:algs} respectively
$$
e_t = \left(C_2+ \frac{C_1}{r^2}-\L r^2\right)^{-1/2}\p_t
\quad\text{and}\quad
\s^2 =\frac{C_2}{r^2 \sin^2 \left(\sqrt{C_2} z\right)}\ .
$$ 
This solution is the metric~\eqref{kLambda}.
\item\underline{Case BC.2: $\k = 0$}\\
Because $\d = \e=0$, we can always find a coordinate system in which
$e_r=\p_r$ and $e_z =\p_z$.
It then follows from~\eqref{eq:algs} and~\eqref{comm:et} that the metric is of the 
form
$$ds^2 = W(r)dt^2 + dr^2 + dz^2 + \frac{1}{\s(r)^2} d\O^2.$$
Hence, it is the product of two Einstein spaces, leading to the product of two 
spheres~\eqref{S2S3}. 
\end{itemize}
\section{Type~$\mathbf{22}$: solutions 
in cases A, B and C}\label{ap:ABandC}
In this appendix, we classify all solutions of the field equations given in
appendix~\ref{veldvgl}, subject to the conditions of case A, B and C of
~\eqref{Pcond2}.
\boldmath
\subsection*{Case A: $\vp_3=\vp_4=0$}
\unboldmath
The Bianchi identities give 
\begin{equation}\label{A:constr}
\k \vp_1 -(\k+\m)\vp_2=0
\qquad\text{and}\qquad
-\l \vp_1 +(\l+\n)\vp_2=0.
\end{equation}
Because we are not interested in the trivial case $\vp_1=\vp_2=0$, it follows
from these equations that $\k\n=\l\m$. We choose our tetrad such that $\n=0$.
At this point, it is necessary to resolve two cases: $\m=0$ and $\m\neq0$.
\begin{itemize}
\item\underline{Case A.1: $\m=0$}\\
From equations~\eqref{eq:alg} and~\eqref{eq:erm}, 
we have $\L=0$. We are not interested in this
case, a discussion can be found in~\cite{0206106}.
\item\underline{Case A.2: $\m\neq0$}\\
In this case, we have $\l=0$. From equation~\eqref{eq:ern} we find $\d=0$, and
from equations~\eqref{eq:alg},~\eqref{eq:ean},~\eqref{eq:eal} 
and~\eqref{A:constr} we obtain
$$\k (\k+\m)=0,\qquad
\vp_1 = \frac{\e}{2}(\k+\m)
\qquad\text{and}\qquad
\vp_2 = \frac{\k}{2}(\e+\m).
$$
Hence we have $\vp_1=0$ or $\vp_2=0$, leading to cases~AC or~AD. The discussion
in appendix~\ref{ap:22s} shows that this leads to the five-sphere or the 
product of two spheres~\eqref{S2S3}.
\end{itemize}
\boldmath
\subsection*{Case B: $\vp_1+\vp_2=0$}
\unboldmath
We choose the tetrad in such a way that $\vp_4=0$.
From the Bianchi equations~\eqref{eq:Bianchi}, we get 
$\m(\vp_1-\vp_3)=0$ and $\n(\vp_1+\vp_3)=0$. 
Hence, there are two possibilities. The first one is $\m\neq 0$ or $\n\neq 0$.
Then we have $\vp_1=\vp_3$ or $\vp_1=-\vp_3$. Therefore the metric has
type~\underline{22}. This case is treated in case~BC of appendix~\ref{ap:22s}.
The second possibility is $\m=\n=0$. Then, it follows from~\eqref{comm:et}
that the manifold is a product manifold 
with metric of the form $ds^2 = dt^2 + dM^2$. $\L$ has to be zero for this metric
to be Einstein.
\boldmath
\subsection*{Case C: $\vp_1^2=\vp_3^2+\vp_4^2$}
\unboldmath
We choose our tetrad such that $\vp_4=0$ and $\vp_3=\vp_1$. A similar analysis as 
the one in~\cite{0206106} shows that we have either case~AC,~BC or 
$$
\l = 0,\ \n=0, \qquad\vp_1 = -\frac{1}{8}(\m+\e)(\k+\m)
\qquad\text{and}\qquad
\vp_2 = \frac{1}{8}(\m+\e)(3\k-2\e+\m).
$$
Inserting these relations into~\eqref{eq:curva} leads to $\L=0$.
\boldmath
\section[Type 22: solutions in case D]{Type 22: solutions in case D, $\vp_2^2=\vp_3^2+\vp_4^2$}
\label{ap:D}
\unboldmath
We choose the tetrad in such a way that $\vp_4=0$ and $\vp_3=-\vp_2$. From the 
Bianchi identities, it follows that $\vp_2=0$ or that $\d=\n$. In the former case,
the discussion in case AD shows that this leads to the five-sphere or the product
of two spheres. We will now proceed with the latter case $\d=\n$. An analysis 
similar to the one in~\cite{0206106} shows
that the problem is split into three cases
\begin{align*}
\text{D.1}\quad &\e=\k\neq 0\quad\text{ and }\quad \vp_1+\vp_2=0\ ,\\
\text{D.2}\quad &\e=0, \quad\k\neq 0\quad \text{ and }\quad \vp_1+3\vp_2=0\ ,\\
\text{D.3}\quad &\e=\k=0\ .
\end{align*}
\boldmath
\subsection*{Case D.1: $\e=\k\neq 0$ and $\vp_1+\vp_2=0$}
\unboldmath
This case is treated in case~BC of appendix~\ref{ap:22s}.
\boldmath
\subsection*{Case D.2: $\e=0$, $\k\neq 0$ and $\vp_1+3\vp_2=0$}
\unboldmath
Inserting the relation $\vp_1+3\vp_2=0$ into the Bianchi equations gives 
$\l=-\n$. From equation~\eqref{eq:alg}, we get 
$\vp_2 = 1/4 (\k\m-\n^2-\L)$. Inserting these results into
equations~\eqref{eq:curvr} and~\eqref{eq:curva} gives\\
\begin{subequations}\label{D2:er}
\begin{minipage}{7.5cm}
\begin{align}
e_r(\m) &=-3\n^2-\m^2+2\k\m-3 \L\\
e_r(\n) &=0\label{D2:ern}\\
e_r(\k) &=\k^2+\k\m
\end{align}
\end{minipage}
\end{subequations}
\begin{subequations}\label{D2:ea}
\begin{minipage}{7.5cm}
\begin{align}
e_z(\m) &=-\n\m\\
e_z(\n) &=-\n^2-\L\label{D2:ean}\\
e_z(\k) &=-\n\k
\end{align}
\end{minipage}
\end{subequations}
\\[2ex]
We choose coordinates in such a way that $e_r = A\p_r$ and $e_z=B\p_z$. 
From the commutation relation~\eqref{comm:erea}, it follows that $B$ depends
only on the coordinate $z$, hence, by a coordinate transformation, we can put
$B=\sqrt{\L}$. The solution of~\eqref{D2:ean} splits into two cases.
\begin{itemize}
\item\underline{$\n$ is not constant}\\
Then we have $\n= \sqrt{\L} \cot z$. We have removed the integration constant by 
a coordinate transformation on $z$. From equations~\eqref{D2:ea}, we have
$$\m = \frac{f(r)}{\sin z} \qquad\text{and}\qquad \k = \frac{g(r)}{\sin z}.$$
From the commutation relations, it follows that we can take 
$e_r = -\frac{r g(r)}{\sin z} \p_r$
without loosing generality. The solution of~\eqref{D2:er} then reads
$$f(r) g(r) = \L - \frac{m}{r^3} \qquad\text{and}\qquad 
g(r)=  \frac{1}{r}\left[ C_1 -\frac{2 m}{r} - \L r^2\right]^{1/2}.$$
This leads to the metric~\eqref{LamkegeldSS}.
\item\underline{$\n$ is constant}\\
In a similar way as above, one can show that this case leads to the black string
in AdS~\eqref{BS}.
\end{itemize}
\boldmath
\subsection*{Case D.3: $\e=\k=0$}
\unboldmath
From~\eqref{eq:alg}, we get $\vp_1=\vp_2-\l\n+\L $. From~\eqref{eq:curvr} 
and~\eqref{eq:curva} we obtain the following equations\\
\begin{minipage}{7.5cm}
\begin{equation}\label{D3:er}
\begin{split}
e_r(\m) &=-\n^2-\m^2+\l\n+4\vp_2-2 \L\\
e_r(\n) &=0\\
e_r(\l) &=0
\end{split}
\end{equation}
\end{minipage}
\begin{minipage}{7.5cm}
\begin{equation}\label{D3:ea}
\begin{split}
e_z(\m) &=-\n\m\\
e_z(\n) &=-\n^2+\l\n-4\vp_2-2\L\\
e_z(\l) &=\l^2+\l\n-4\vp_2
\end{split}
\end{equation}
\end{minipage}\\[2ex]
We choose
coordinates in such a way that $e_r = A\p_r$ and $e_z = B\p_z$. From the
commutation relation~\eqref{comm:erea}, it follows that we can do a coordinate
transformation to make $A$ and $B$ both depend only on $z$. The solution splits 
into three cases
\begin{itemize}
\item\underline{$\m=0$}\\
From~\eqref{eq:algs} and~\eqref{comm:et}, it follows that the metric can be 
written as
$$ds^2 = \frac{1}{A(z)^2} (dt^2 + dr^2)+ \frac{1}{B(z)^2}dz^2 +
\frac{1}{\s(z)^2} d\O^2.$$
This metric belongs to the class~\eqref{sphere-constant}, where the 
factor is a two-plane. We have not been able to find an 
analytical form of the general solution of~\eqref{D3:er} and~\eqref{D3:ea}.
\item\underline{$\m\neq0$ and $e_r(\mu)=0$}\\
In this cases, the metric can be written as
$$ds^2 = \frac{1}{A(z)^2} (dr^2 + e^{ - 2 C r} dt^2)+ \frac{1}{B(z)^2}dz^2 +
\frac{1}{\s(z)^2} d\O^2.$$
This metric belongs to the class~\eqref{sphere-constant}, where the second 
factor is the two-sphere if $C^2 <  0$ and the two-dimensional hyperbolic space if
$C^2 > 0$. We have not been able to find an 
analytical form of the general solution of~\eqref{D3:er} and~\eqref{D3:ea}.
\item\underline{$\m\neq0$ and $e_r(\mu)\neq0$}\\
This case leads to the same metrics as above.
\end{itemize}

\end{document}